\documentclass[twocolumn,floatfix,showpacs,eqsecnum]{revtex4}
\usepackage{graphicx}
\usepackage{amsmath}
\usepackage{bm}
\begin{document}

\title{Nonalgebraic length dependence of transmission through a chain of barriers with a L\'{e}vy spacing distribution}

\author{C. W. J. Beenakker, C. W. Groth, and A. R. Akhmerov}
\affiliation{Instituut-Lorentz, Universiteit Leiden, P.O. Box 9506, 2300 RA Leiden, The Netherlands}

\date{November 2008}
\begin{abstract}
The recent realization of a ``L\'{e}vy glass'' (a three-dimensional optical material with a L\'{e}vy distribution of scattering lengths) has motivated us to analyze its one-dimensional analogue: A linear chain of barriers with independent spacings $s$ that are L\'{e}vy distributed: $p(s)\propto s^{-1-\alpha}$ for $s\rightarrow\infty$. The average spacing diverges for $0<\alpha<1$. A random walk along such a sparse chain is not a L\'{e}vy walk because of the strong correlations of subsequent step sizes. We calculate all moments of conductance (or transmission), in the regime of incoherent sequential tunneling through the barriers. The average transmission from one barrier to a point at a distance $L$ scales as $L^{-\alpha}\ln L$ for $0<\alpha<1$. The corresponding electronic shot noise has a Fano factor ($\propto$ average noise power / average conductance) that approaches $1/3$ very slowly, with $1/\ln L$ corrections.
\end{abstract}
\maketitle

\section{Introduction}

In a recent publication \cite{Bar08}, Barthelemy, Bertolotti, and Wiersma have reported on the fabrication of an unusual random optical medium which they have called a \textit{L\'{e}vy glass.} It consists of a random packing of glass microspheres having a L\'{e}vy distribution of diameters. The space between the spheres is filled with strongly scattering nanoparticles. A photon trajectory therefore consists of ballistic segments of length $s$ through spherical regions, connected by isotropic scattering events. A L\'{e}vy distribution is characterized by a slowly decaying tail, $p(s)\propto 1/s^{1+\alpha}$ for $s\rightarrow\infty$, with $0<\alpha<2$, such that the second moment (and for $\alpha<1$ also the first moment) diverges. The transmission of light through the L\'{e}vy glass was analyzed \cite{Bar08} in terms of a L\'{e}vy walk \cite{Man83,Shl95,Met00} for photons.

Because the randomness in the L\'{e}vy glass is frozen in time (``quenched'' disorder), correlations exist between subsequent scattering events. Backscattering after a large step is likely to result in another large step. This is different from a L\'{e}vy walk, where subsequent steps are independently drawn from the L\'{e}vy distribution (``annealed'' disorder). Numerical \cite{Kut98} and analytical \cite{Sch02} theories indicate that the difference between quenched and annealed disorder can be captured (at least approximately) by a renormalization of the L\'{e}vy walk exponent --- from the annealed value $\alpha$ to the quenched value $\alpha'=\alpha+(2/d)\max(0,\alpha-d)$ in $d$ dimensions. Qualitatively speaking, the correlations in a L\'{e}vy glass slow down the diffusion relative to what is expected for a L\'{e}vy walk, and the effect is the stronger the lower the dimension.

To analyze the effect of such correlations in a quantitative manner, we consider in this paper the one-dimensional analogue of a L\'{e}vy glass, which is a linear chain of barriers with independently L\'{e}vy distributed spacings $s$. Such a system might be produced artificially, along the lines of Ref.\ \cite{Bar08}, or it might arise naturally in a porous medium \cite{Lev97} or in a nanowire \cite{Koh04}. Earlier studies of this system \cite{Hav86,Che92,Bar00,note1,Boo07} have compared the dynamical properties with those of a L\'{e}vy walk. In particular, Barkai, Fleurov, and Klafter \cite{Bar00} found a superdiffusive mean-square displacement as a function of time [$\langle x^{2}(t)\rangle\propto t^{\gamma}$ with $\gamma>1$] --- reminiscent of a L\'{e}vy walk (where $\gamma=3-\alpha$). No precise correspondence to a L\'{e}vy walk is to be expected in one dimension, because subsequent step lengths are highly correlated: Backscattering after a step of length $s$ to the right results in the same step length $s$ to the left.

The simplicity of one-dimensional dynamics allows for an exact solution of the static transmission statistics, without having to assume a L\'{e}vy walk. We present such a calculation here, and find significant differences with the $L^{-\alpha/2}$ scaling of the average transmission expected \cite{Dav97,Lar98,Bul01} for a L\'{e}vy walk (annealed disorder) through a system of length $L$. If the length of the system is measured from the first barrier, we find for the case of quenched disorder an average transmission $\langle T\rangle\propto L^{-\alpha}\ln L$ for $0<\alpha<1$ and $\langle T\rangle\propto L^{-1}$ for $\alpha>1$. Note that the nonalgebraic length dependence for $0<\alpha<1$ goes beyond what can be captured by a renormalization of $\alpha$.

In the electronic context the average conductance $\langle G\rangle$ is proportional to $\langle T\rangle$, in view of the Landauer formula. In that context it is also of interest to study the shot noise power $S$, which quantifies the time dependent fluctuations of the current due to the granularity of the electron charge. We calculate the Fano factor $F\propto\langle S\rangle/\langle G\rangle$, and find that $F$ approaches the value $1/3$ characteristic of normal diffusion \cite{Bee92,Nag92} with increasing $L$ --- but with relatively large corrections that decay only as $1/\ln L$ for $0<\alpha<1$.

\section{Formulation of the problem}

\begin{figure}[tb]
\centerline{\includegraphics[width=0.8\linewidth]{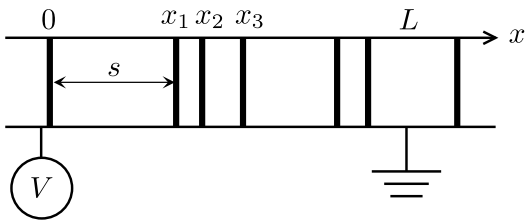}}
\caption{\label{fig_layout}
Linear chain of randomly spaced tunnel barriers. We study the statistics of conductance (or transmission) over a length $L$ for a L\'{e}vy distribution of spacings $p(s)$.
}
\end{figure}

We consider a linear chain of tunnel barriers, see Fig.\ \ref{fig_layout}, with a distribution of spacings $p(s)$ that decays for large $s$ as $1/s^{\alpha+1}$. A normalizable distribution requires $\alpha>0$. For $0<\alpha<1$ the mean spacing is infinite. We take for each barrier the same mode-independent transmission probability $\Gamma\ll 1$ (no ballistic transmission). The corresponding tunnel resistance is $r=(h/e^{2})(N\Gamma)^{-1}$, with $N$ the number of transverse modes. In the electronic context we require $r\ll h/e^{2}$, so that the Coulomb blockade of single-electron tunneling can be ignored. 

We work in the regime of incoherent sequential tunneling (no resonant tunneling). This regime can be reached for $N\gg 1$ as a result of intermode scattering, or it can be reached even for small $N$ as a result of a short phase coherence length. For sequential tunneling the resistance $R$ of $n$ barriers in series is just the series resistance $nr$ [corresponding to a transmission probability $T=(n\Gamma)^{-1}$]. We measure this resistance 
\begin{equation}
R(L)=r\sum_{n}\theta(x_{n})\theta(L-x_{n}) \label{R1def}
\end{equation}
between one contact at $x=0$ and a second contact at $x=L>0$. The numbers $x_{n}$ indicate the coordinates of the tunnel barriers and $\theta(x)$ is the step function [$\theta(x)=1$ if $x>0$ and $\theta(x)=0$ if $x<0$].

Without further restrictions the statistics of the conductance would be dominated by ballistic realizations, that have not a single tunnel barrier in the interval $(0,L)$. The reason, discussed in Ref.\ \cite{Bar00}, is that the average distance between a randomly chosen point along the chain and the nearest tunnel barrier diverges for any $0<\alpha<2$ (so even if the mean spacing between the barriers is finite). To eliminate ballistic transmission, we assume that one tunnel barrier is kept fixed at $x_{0}=0^{+}$. (This barrier thus contributes $r$ to the resistance.) If we order the coordinates such that $x_{n}<x_{n+1}$, we have
\begin{equation}
R(L)=r+r\sum_{n=1}^{\infty}\theta(x_{n})\theta(L-x_{n}). \label{R2def}
\end{equation}

We seek the scaling with $L$ in the limit $L\rightarrow\infty$ of the negative moments
$\langle R(L)^{p}\rangle$ ($p=-1,-2,-3,\ldots$) of the resistance. This information will give us the scaling of the positive moments of the conductance $G=R^{-1}$ and transmission $T=(h/Ne^{2})R^{-1}$. It will also give us the average of the shot noise power $S$, which for an arbitrary number of identical tunnel barriers in series is determined by the formula \cite{Jon95}
\begin{equation}
S=\frac{2}{3}e|V|r^{-1}[(R/r)^{-1}+2(R/r)^{-3}],\label{Sdef}
\end{equation}
where $V$ is the applied voltage. From $\langle S\rangle$ and $\langle G\rangle$ we obtain the Fano factor $F$, defined by
\begin{equation}
F=\frac{\langle S\rangle}{2e|V|\langle G\rangle}.\label{Fanodef}
\end{equation}

\section{Arbitrary moments}
\label{moments}

The general expression for moments of the resistance is
\begin{equation}
\langle R(L)^{p}\rangle=r^{p}\left\langle\left(1+\sum_{n=1}^{\infty}\theta(x_{n})\theta(L-x_{n})\right)^{p}\right\rangle,\label{Rpdef}
\end{equation}
where the brackets $\langle\cdots\rangle$ indicate the average over the spacings,
\begin{equation}
\langle\cdots\rangle=\prod_{n=1}^{\infty}\int_{-\infty}^{\infty}dx_{n}\,p(x_{n}-x_{n-1})\cdots,\label{averagedef}
\end{equation}
with the definitions $x_{0}=0$ and $p(s)=0$ for $s<0$. We work out the average,
\begin{align}
\langle R(L)^{p}\rangle={}&r^{p}\sum_{n=1}^{\infty}n^{p}\left(\prod_{i=1}^{n}\int_{-\infty}^{\infty}ds_{i}\,p(s_{i})\right)\nonumber\\
&\times\theta\left(\sum_{i=1}^{n}s_{i}-L\right)\theta\left(L-\sum_{i=1}^{n-1}s_{i}\right).\label{Rpworkedout}
\end{align}

It is more convenient to evaluate the derivative with respect to $L$ of Eq.\ \eqref{Rpworkedout}, which takes the form of a multiple convolution of the spacing distribution \cite{note2},
\begin{widetext}
\begin{equation}
\frac{d}{dL}\langle R^{p}\rangle=r^{p}(2^{p}-1)p(L)+r^{p}\sum_{n=2}^{\infty}[(n+1)^{p}-n^{p}]\int_{-\infty}^{\infty}dx_{n-1}\cdots\int_{-\infty}^{\infty}dx_{1}
\,p(L-x_{n-1})p(x_{n-1}-x_{n-2})\cdots p(x_{2}-x_{1})p(x_{1}).
\label{Rpderivative}
\end{equation}
\end{widetext}
In terms of the Fourier (or Laplace) transform
\begin{equation}
f(\xi)=\int_{0}^{\infty}ds\,e^{i\xi s}p(s),\label{fxidef}
\end{equation}
the series \eqref{Rpderivative} can be summed up,
\begin{align}
&\frac{d}{dL}\langle R^{p}\rangle=\frac{r^{p}}{2\pi}\int_{-\infty+i0^{+}}^{\infty+i0^{+}}d\xi\,e^{-i\xi L}\sum_{n=1}^{\infty}[(n+1)^{p}-n^{p}]f(\xi)^{n}\nonumber\\
&\qquad=\frac{r^{p}}{2\pi}\int_{-\infty+i0^{+}}^{\infty+i0^{+}}d\xi\,e^{-i\xi L}\frac{1-f(\xi)}{f(\xi)}\,{\rm Li}_{-p}[f(\xi)].\label{sum}
\end{align}
The function $\mbox{Li}(x)$ is the polylogarithm. The imaginary infinitesimal $i0^{+}$ added to $\xi$ regularizes the singularity of the integrand at $\xi=0$. For negative $p$ this singularity is integrable, and the integral \eqref{sum} may be rewritten as an integral over the positive real axis,
\begin{equation}
\frac{d}{dL}\langle R^{p}\rangle=\frac{r^{p}}{\pi}\mbox{Re}\,\int_{0}^{\infty}d\xi\,e^{-i\xi L}\frac{1-f(\xi)}{f(\xi)}\,{\rm Li}_{-p}[f(\xi)].\label{sumreal}
\end{equation}

\section{Scaling with length}
\label{scaling}

\subsection{Asymptotic expansions}
\label{asymptotico}

In the limit $L\rightarrow\infty$ the integral over $\xi$ in Eq.\ \eqref{sumreal} is governed by the $\xi\rightarrow 0$ limit of the Fourier transformed spacing distribution. Because $p(s)$ is normalized to unity one has $f(0)=1$, while the large-$s$ scaling $p(s)\propto 1/s^{\alpha+1}$ implies
\begin{equation}
\lim_{\xi\rightarrow 0}f(\xi)=\left\{\begin{array}{l}
1+c_{\alpha}(s_{0}\xi)^{\alpha},\;\;0<\alpha<1,\\
1+i\bar{s}\xi+c_{\alpha}(s_{0}\xi)^{\alpha},\;\;1<\alpha<2.
\end{array}\right.
\label{fxilimit}
\end{equation}
The characteristic length $s_{0}>0$, the mean spacing $\bar{s}$, as well as the numerical coefficient $c_{\alpha}$ are determined by the specific form of the spacing distribution.

The limiting behavior of the polylogarithm is governed by
\begin{align}
&{\rm Li}_{1}(1+\epsilon)=-\ln(-\epsilon),\label{Li1limit}\\
&\lim_{\epsilon\rightarrow 0}{\rm Li}_{2}(1+\epsilon)=\zeta(2)-\epsilon\ln(-\epsilon),\label{Li2limit}\\
&\lim_{\epsilon\rightarrow 0}{\rm Li}_{n}(1+\epsilon)=\zeta(n)+\zeta(n-1)\varepsilon,\;\;n=3,4,\ldots\label{Lin3limit}
\end{align}
In combination with Eq.\ \eqref{fxilimit} we find, for $0<\alpha<1$, the following expansions of the integrand in Eq.\ \eqref{sumreal}:
\begin{align}
\lim_{\xi\rightarrow 0}\frac{1-f}{f}\,{\rm Li}_{-p}(f)={}&c_{\alpha}(s_{0}\xi)^{\alpha}\ln[-c_{\alpha}(s_{0}\xi)^{\alpha}],\nonumber\\
&\mbox{if}\;\;p=-1,\label{Lipmin1}\\
\lim_{\xi\rightarrow 0}\frac{1-f}{f}\,{\rm Li}_{-p}(f)={}&-\zeta(-p)c_{\alpha}(s_{0}\xi)^{\alpha},\nonumber\\
&p=-2,-3\ldots\label{Lipmin2}
\end{align}
For $1<\alpha<2$ we should replace $c_{\alpha}(s_{0}\xi)^{\alpha}$ by $i\bar{s}\xi+c_{\alpha}(s_{0}\xi)^{\alpha}$.

\subsection{Results}
\label{results}

We substitute the expansions \eqref{Lipmin1} and \eqref{Lipmin2} into Eq.\ \eqref{sumreal}, and obtain the large-$L$ scaling of the moments of conductance with the help of the following Fourier integrals ($L>0$, $\alpha>-1$):
\begin{align}
&\int_{0}^{\infty}d\xi\,e^{-i\xi L}\xi^{\alpha}\ln\xi=i\Gamma(1+\alpha)e^{-i\pi\alpha/2}L^{-1-\alpha}\nonumber\\
&\qquad\qquad\times(\ln L+i\pi/2+\gamma_{E}-H_{\alpha}),\label{int0}\\
&\int_{0}^{\infty}d\xi\,e^{-i\xi L}\xi^{\alpha}=-i\Gamma(1+\alpha)e^{-i\pi\alpha/2}L^{-1-\alpha},\label{int2}\\
&\mbox{Re}\,\int_{0}^{\infty}d\xi\,e^{-i\xi L}i\xi=0,\label{int00}\\
&\mbox{Re}\,\int_{0}^{\infty}d\xi\,e^{-i\xi L}i\xi\ln\xi=-\tfrac{1}{2}\pi L^{-2}.\label{int11}
\end{align}
Here $\gamma_{E}$ is Euler's constant and $H_{\alpha}$ is the harmonic number. The resulting scaling laws are listed in Table \ref{scalingtable}.

\begin{table}
\caption{\label{scalingtable} Scaling with $L$ of moments of conductance (or, equivalently, transmission).}
\begin{ruledtabular}
\begin{tabular}{l|ll}
&$0<\alpha<1$&$1<\alpha<2$\\
\hline
$\langle R^{-1}\rangle\equiv\langle G\rangle$&$L^{-\alpha}\ln L$&$L^{-1}$\\
$\langle R^{p}\rangle\equiv\langle G^{-p}\rangle$, $p=-2,-3,\ldots$&$L^{-\alpha}$&$L^{-\alpha}$
\end{tabular}
\end{ruledtabular}
\end{table}

Two physical consequences of these scaling laws are:
\begin{itemize}
\item
The Fano factor \eqref{Fanodef} approaches $1/3$ in the limit $L\rightarrow\infty$, regardless of the value of $\alpha$, but for $0<\alpha<1$ the approach is very slow: $F-1/3\propto 1/\ln L$. For $1<\alpha<2$ the approach is faster but still sublinear, $F-1/3\propto 1/L^{\alpha-1}$.
\item
The root-mean-square fluctuations $\mbox{rms}\,G=\sqrt{\langle G^{2}\rangle-\langle G\rangle^{2}}$ of the conductance become much larger than the average conductance for large $L$, scaling as $\mbox{rms}\,G/\langle G\rangle\propto L^{\alpha/2}/\ln L$ for $0<\alpha<1$ and as $\mbox{rms}\,G/\langle G\rangle\propto L^{1-\alpha/2}$ for $1<\alpha<2$.
\end{itemize}

\section{Numerical test}

To test the scaling derived in the previous sections, in particular to see how rapidly the asymptotic $L$-dependence is reached with increasing $L$, we have numerically generated a large number of random chains of tunnel barriers and calculated moments of conductance and the Fano factor from Eqs.\ \eqref{R2def}--\eqref{Fanodef}.

For the spacing distribution in this numerical calculation we took the L\'{e}vy stable distribution \cite{Man94} for $\alpha=1/2$,
\begin{equation}
p_{1/2}(s)=(s_{0}/2\pi)^{1/2}s^{-3/2}e^{-s_{0}/2s}.\label{pstabledef}
\end{equation}
Its Fourier transform is
\begin{equation}
f_{1/2}(\xi)=\exp(-\sqrt{-2is_{0}\xi})\Rightarrow c_{1/2}=i-1. \label{fstabledef}
\end{equation}

Inserting the numerical coefficients, the large-$L$ scaling of conductance moments for the distribution \eqref{pstabledef} is
\begin{align}
&\lim_{L\rightarrow\infty}\langle G\rangle=\frac{1}{r}(2\pi L/s_{0})^{-1/2}[\ln(2L/s_{0})+\gamma_{E}],\label{GLevy}\\
&\lim_{L\rightarrow\infty}\langle G^{p}\rangle=2\zeta(p)\frac{1}{r^{p}}(2\pi L/s_{0})^{-1/2},\;\;p\geq 2.\label{GnLevy}
\end{align}
The resulting scaling of the conductance fluctuations and Fano factor is
\begin{align}
&\left(\frac{\mbox{rms}\,G}{\langle G\rangle}\right)^{2}\equiv\frac{\langle G^{2}\rangle}{\langle G\rangle^{2}}-1\approx\frac{(\pi^{2}/3)(2\pi L/s_{0})^{1/2}}{[\ln(2L/s_{0})+\gamma_{E}]^{2}}-1,\label{rmsLevy}\\
&F\approx\frac{1}{3}+\frac{(4/3)\zeta(3)}{\ln(2L/s_{0})+\gamma_{E}}.\label{FLevy}
\end{align}

\begin{figure}[tb]
\centerline{\includegraphics[width=0.8\linewidth]{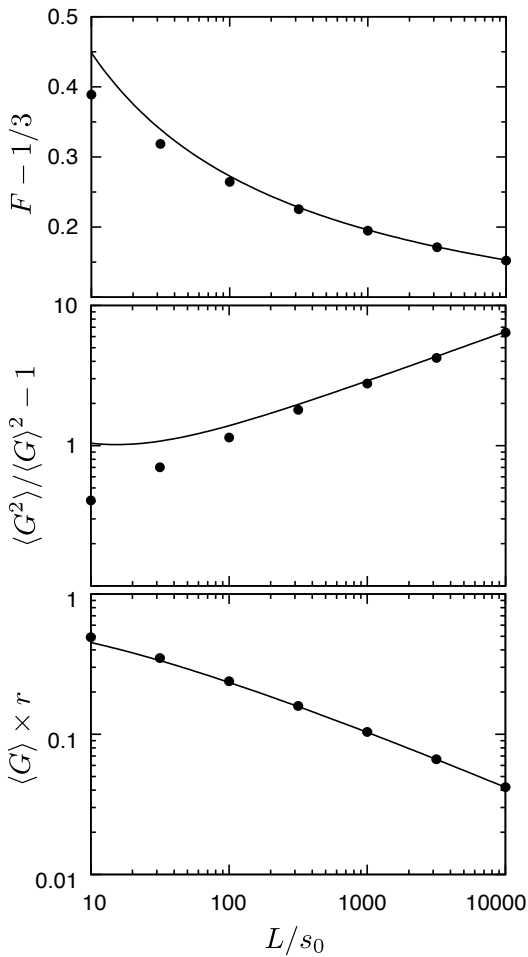}}
\caption{\label{fig_combined}
Scaling of the average conductance (bottom panel), the variance of the conductance (middle panel), and the Fano factor (top panel), for a chain of tunnel barriers with spacings distributed according to the $\alpha=1/2$ L\'{e}vy stable distribution \eqref{pstabledef}. The data points are calculated numerically, by averaging over a large number of random chains of tunnel barriers. The solid curves are the analytical results \eqref{GLevy}--\eqref{FLevy} of the asymptotic analysis in the $L\rightarrow\infty$ limit.
}
\end{figure}

In Fig.\ \ref{fig_combined} we compare these analytical large-$L$ formulas with the numerical data. The average conductance converges quite rapidly to the scaling \eqref{GLevy}, while the convergence for higher moments (which determine the conductance fluctuations and Fano factor) requires somewhat larger systems. We clearly see in Fig.\ \ref{fig_combined} the relative growth of the conductance fluctuations with increasing system size and the slow decay of the Fano factor towards the diffusive $1/3$ limit.

\section{Conclusion and outlook}

In conclusion, we have analyzed the statistics of transmission through a sparse chain of tunnel barriers. The average spacing of the barriers diverges for a L\'{e}vy spacing distribution $p(s)\propto 1/s^{1+\alpha}$ with $0<\alpha<1$. This causes an unusual scaling with system length $L$ (measured from the first tunnel barrier) of the moments of transmission or conductance, as summarized in Table \ref{scalingtable}. A logarithmic correction to the power law scaling appears for the first moment. Higher moments of conductance all scale with the same power law, differing only in the numerical prefactor. As a consequence, sample-to-sample fluctuations of the transmission become larger than the average with increasing $L$.

This theoretical study of a one-dimensional ``L\'{e}vy glass'' was motivated by a recent optical experiment on its three-dimensional analogue \cite{Bar08}. The simplicity of a one-dimensional geometry has allowed us to account exactly for the correlations between subsequent step lengths, which distinguish the random walk through the sparse chain of barriers from a L\'{e}vy walk. We surmise that step length correlations will play a role in two and three dimensional sparse arrays as well, complicating a direct application of the theory of L\'{e}vy walks to the experiment. This is one line of investigation for the future.

A second line of investigation is the effect of wave interference on the transmission of electrons or photons through a sparse chain of tunnel barriers. Here we have considered the regime of incoherent sequential transmission, appropriate for a multi-mode chain with mode-mixing or for a single-mode chain with a short coherence length. The opposite, phase coherent regime was studied in Ref.\ \cite{Boo07}. In a single-mode and phase coherent chain interference can lead to localization, producing an exponential decay of transmission. An investigation of localization in this system is of particular interest because the sparse chain belongs to the class of disordered systems with long-range disorder, to which the usual scaling theory of Anderson localization does not apply \cite{Rus01}.

A third line of investigation concerns the question ``what is the shot noise of anomalous diffusion''? Anomalous diffusion \cite{Met00} is characterized by a mean square displacement $\langle x^{2}\rangle\propto t^{\gamma}$ with $0<\gamma<1$ (subdiffusion) or $\gamma>1$ (superdiffusion). The shot noise for normal diffusion ($\gamma=1$) has Fano factor $1/3$ \cite{Bee92,Nag92}, and Ref.\ \cite{Gro08} concluded that subdiffusion on a fractal also produces $F=1/3$. Here we found a convergence, albeit a logarithmically slow convergence, to the same $1/3$ Fano factor for a particular system with superdiffusive dynamics. We conjecture that $F=1/3$ in the entire subballistic regime $0<\gamma<2$, with deviations appearing in the ballistic limit $\gamma\rightarrow 2$ --- but we do not have a general theory to support this conjecture.  

\acknowledgments
We have benefited from discussions with R. Metzler and J. Tworzyd{\l}o. This research was supported by the Dutch Science Foundation NWO/FOM.


\begin{thebibliography}{99}
\bibitem{Bar08} P. Barthelemy , J. Bertolotti, and D. S. Wiersma, Nature \textbf{453}, 495 (2008).
\bibitem{Man83} B. B. Mandelbrot, \textit{The Fractal Geometry of Nature} (Freeman, New York, 1983). 
\bibitem{Shl95} M. Shlesinger, G. Zaslavsky, and U. Frisch, editors, \textit{L\'{e}vy Flights and Related Topics in Physics} (Springer, Berlin, 1995). 
\bibitem{Met00} R. Metzler and J. Klafter, Phys. Rep. \textbf{339}, 1 (2000).
\bibitem{Kut98} R. Kutner and Ph.\ Maass, J. Phys. A \textbf{31}, 2603 (1998).
\bibitem{Sch02} M. Schulz, Phys. Lett. A \textbf{298}, 105 (2002); M. Schulz and P. Reineker, Chem. Phys. \textbf{284}, 331 (2002).
\bibitem{Lev97}  P. Levitz, Europhys. Lett. \textbf{39}, 6593 (1997).
\bibitem{Koh04} H. Kohno and H. Yoshida, Solid State Comm. \textbf{132}, 59 (2004).
\bibitem{Hav86} S. Havlin, A. Bunde, and H. E. Stanley, Phys. Rev. B \textbf{34}, 445 (1986). 
\bibitem{Che92} K. W. Cheung, K. W. Yu, and P. M. Hui, Phys. Rev. B \textbf{45}, 456 (1992).
\bibitem{Bar00} E. Barkai, V. Fleurov, and J. Klafter, Phys. Rev. E \textbf{61} 1164 (2000).
\bibitem{note1} The authors of Ref.\ \cite{Bar00} calculate a lower bound to the mean square displacement, with the result $\langle x^{2}\rangle\geq t^{\min(2,3-\alpha)}$ if the initial position of the particle is randomly chosen along the chain (so superdiffusion for any $0<\alpha<2$). If the particle starts at a barrier (which corresponds to the situation we consider in the present work), the result is $\langle x^{2}\rangle\geq t^{2-\alpha}$ (so superdiffusion for $0<\alpha<1$). Earlier papers \cite{Hav86,Che92} gave different results for the mean square displacement, but a direct comparison is problematic because those papers did not notice the dependence on the starting position.
\bibitem{Boo07} D. Boos\'{e} and J. M. Luck, J. Phys. A \textbf{40}, 14045 (2007).
\bibitem{Dav97} A. Davis and A. Marshak, in \textit{Fractal Frontiers}, edited by M. 
M. Novak and T. G. Dewey (World Scientific, 1997).
\bibitem{Lar98} H. Larralde, F. Leyvraz, G. Martinez-Mekler, R. Rechtman, and S. Ruffo, Phys. Rev. E \textbf{58}, 4254 (1998).
\bibitem{Bul01} S. V. Buldyrev, S. Havlin, A. Ya. Kazakov, M. G. E. da Luz, E. P. Raposo, H. E. Stanley, and G. M. Viswanathan, Phys. Rev. E \textbf{64}, 041108 (2001); S. V. Buldyrev et al., Physica A \textbf{302}, 148 (2001).
\bibitem{Bee92} C. W. J. Beenakker and M. B\"{u}ttiker, Phys. Rev. B \textbf{46}, 1889 (1992).
\bibitem{Nag92} K. E. Nagaev, Phys. Lett. A \textbf{169}, 103 (1992).
\bibitem{Jon95} M. J. M. de Jong and C. W. J. Beenakker, Phys. Rev. B \textbf{51}, 16867 (1995).
\bibitem{note2} We cannot directly take the derivative of Eq.\ \eqref{Rpdef}, because that would lead (for $p\neq 1$) to an undefined product of $\theta(L-x)$ and $\delta(L-x)$. No such complication arises if we take the derivative of Eq.\ \eqref{Rpworkedout}.
\bibitem{Man94} For efficient algorithms to generate random variables with a L\'{e}vy stable distribution, see J. M. Chambers, C. L. Mallows, and B. W. Stuck, J. Am.
Stat. Ass. \textbf{71}, 340 (1976); R. N. Mantegna, Phys. Rev. E \textbf{49}, 4677 (1994).
\bibitem{Rus01} S. Russ, J. W. Kantelhardt, A. Bunde, and S. Havlin, Phys. Rev. B \textbf{64}, 134209 (2001).
\bibitem{Gro08} C. W. Groth, J. Tworzyd{\l}o, and C. W. J. Beenakker, Phys. Rev. Lett. \textbf{100}, 176804 (2008).
\end{thebibliography}
\end{document}